# Temperature dependence of elastocaloric effect in natural rubber

Zhongjian Xie[1], Gael Sebald, Daniel Guyomar

Université de Lyon, INSA-Lyon, LGEF EA682, 8 rue de la Physique, 69621 Villeurbanne, France

## Abstract

The temperature dependence of the elastocaloric (eC) effect in natural rubber (NR) is studied adiabatically and isothermally. A broad temperature span for eC effect from 0 $^{o}$C to 49 $^{o}$C is observed. The maximum adiabatic temperature change ($\Delta T$) is 12 K at strain of 6 and occurs at 10 $^{o}$C. These behaviors can be predicted by the temperature dependence of strain-induced crystallization (SIC) and temperature-induced crystallization (TIC). In isothermal condition, the deduced $\Delta T$ from Clausius-Clapeyron factor can agree with the direct measurement at different temperatures. The eC performance of NR is compared with shape memory alloys (SMAs). The potential of NR for near room temperature and large-scale cooing application is proved. This will open the SIC research of NR towards eC cooling direction.

Current dominant cooling technology is the vapor-compression technology, which is based on the hazardous gas. Solid-state cooling technology based on caloric effect can be an alternative. Caloric effect refers to the adiabatic temperature change or isothermal entropy change stimulated by some fields. According to the stimulus field, caloric effect is divided into magnetocaloric (MC), electrocaloric (EC), barocaloric (BC) and elastocaloric (eC) effect. Although this field is developed from the intention of finding environmentally friendly technology, some caloric effects still show some environmental problems. For MC effect, the needed magnets and the MC material are mainly based on the rare-earth elements (REEs) [1,2], whose production is detrimental to environment [3]. The arsenic (As)-based MC material [4] and the lead-based EC materials [5,6] possess high caloric performance but they are toxic. Some other caloric materials are environmentally friendly and show a high caloric performance but with a high cost, like the widely researched PVDF-based polymers [7,8]. For eC and BC effect, the promising materials are shape memory alloys (SMAs), but they need a large stress (several hundreds of MPa) [9–16], which is not practicable [17]. Thus, alternative caloric materials which are environmentally friendly, cost-effective, of high caloric performance and practicable need to be found.

Natural rubber (NR) is one soft eC material. It is environmentally friendly, recyclable and non-toxic. It has the lowest cost among the materials with giant caloric effects [15]. The low stress (several MPa) of NR makes it easier to be manipulated than hard materials (SMAs). The eC effect of NR was found by Gough and further investigated by Joule (the Gough-Joule effect) [18]. It is the oldest known caloric material, but it is only considered for cooling application recently [19–21].

For eC effect of NR, elasticity and eC temperature change ($\Delta T$) are two basic quantities. The elasticity of NR is mainly related to the strain-induced crystallization/crystallite (SIC) [22,23]. The $\Delta T$ is mainly from the latent heat of SIC [24] and a giant eC effect is observed ($\Delta T \sim$10 K) [24,25]. Considering the application of caloric material for a cooling device, the fatigue life plays a critical role. The fatigue life of NR can be up to $10^7$ cycles for strain amplitude of 200% [26] due to the excellent crack growth resistance of SIC [27,28]. Thus, the research on eC effect of NR should take full account of the SIC theory due to its main contribution for elasticity, $\Delta T$ and fatigue life. Due to the developed wide-angle X-ray diffraction (WAXD) by using synchrotron radiation [29], the SIC research is blossoming. It can help to understand the eC effect deeply and promote its research.

---

[1] zhongjian.xie@insa-lyon.fr
   gael.sebald@insa-lyon.fr

Cooling device works near room temperature. It normally consists of a hot end and a cold end, like the active magnetic regenerator (AMR) [30]. The caloric material needs transfer the heat from the cold end to hot end. Thus, it is important to study the temperature dependence of the caloric material in the working temperature span between hot and cold ends near room temperature. The temperature dependence of SMAs has been widely researched [12,31–34]. A large temperature span was observed in Cu-Zn-Al [33], Ni-Ti [11] and Fe-31.2Pd [35]. For the eC effect in NR, the responsible SIC mechanism shows a temperature dependence [36,37], which may correspondingly predict its temperature dependent behavior.

In this letter, the temperature dependence of eC effect of NR is studied adiabatically and isothermally. A large temperature span near room temperature is shown. The responsible SIC mechanism is discussed. The potential of NR for near room temperature cooling application is proved. Accordingly, it will open the new research direction for SIC.

The NR material was bought from Xinyinte Rubber Products Co., Ltd. (China). It has a network chain density ($N$) value of $1.3 \times 10^{-4} mol \cdot cm^{-3}$, which is calculated from the initial regime of the stress-strain curve. The cross section area of the NR film is 20mm x 100μm. The engineering strain $\varepsilon = (l - l_0)/l_0$ is the ratio of the elongated length $l - l_0$ to the initial length $l_0$.

The tensile tests used an Ironless Linear Motor (XM-550, Newport, New-York) and the stress was measured using a force sensor (ELPF-T2M-250N, Measurement Specialities, Paris). The temperature change was measured using a fine thermocouple in contact with the sample. To measure the temperature dependence of eC effect of NR, the experimental setup was adapted in order to place the NR sample inside a temperature controlled oven, while leaving motor and force sensor outside the oven in order to avoid the disturbance. The temperature dependence was measured adiabatically and isothermally.

For the adiabatic condition, a typical measurement of stress and temperature change was shown in Fig. 1. Firstly, the NR sample was stretched rapidly (strain rate of 20 $s^{-1}$) to a given strain and temperature increased. Secondly, the stretched state was kept for 30 s and temperature decreased due to the heat transfer with outer medium. Finally, the sample retracted and its temperature decreased. This step stretching was performed at different temperatures (0 °C, 10 °C, 17 °C, 27 °C, 39 °C and 49 °C) and at different strains from 1 to 6. The results were shown in Fig. 2 and Fig. 3.

In Fig. 1 (b), the temperature decrease in retraction is larger than temperature increase in extension. This is due to the different kinetics of SIC and melting [21,38]. Temperature decrease in retraction is much closer to the equilibrium state. Thus, it is considered as the adiabatic temperature change and shown in the following figures and texts.

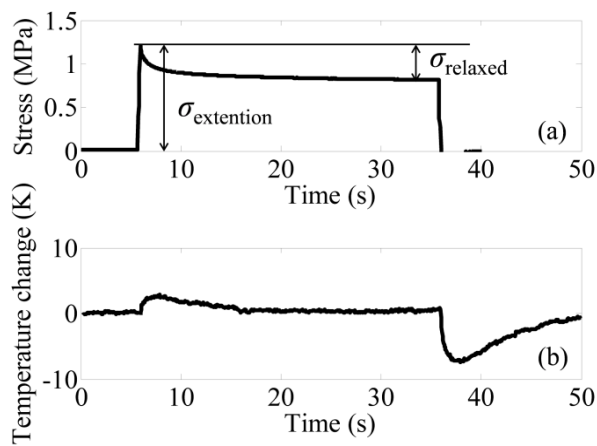

Fig. 1 Typical (a) engineering stress - strain and (b) temperature change - time signal of NR at strain of 6 at 27 °C.

In isothermal condition, the stresses at different static temperatures (identical to the adiabatic condition) were measured in extension and retraction. The strain rate was 0.02 s$^{-1}$. During these measurements, the temperature changes were less than ±1 K and it was considered to be isothermal. The results were shown in Fig. 4. In both adiabatic and isothermal conditions, pre-strain of 1 was applied. The static stress at pre-strain of 1 was set to be zero.

In adiabatic condition, the extension stresses (indicated in Fig. 1 (a)) at different temperatures are compared in Fig. 2 (a). There is no obvious difference (compared with those in isothermal condition in Fig. 4 (a)). The mechanical behavior of NR is mainly attributed to SIC [39] and SIC is sensitive to the temperature [36,37]. Thus, the mechanical behavior of NR is temperature dependent [23,36]. But this phenomenon can be observed only at very low strain rate (10$^{-3}$ s$^{-1}$) [36], because the SIC needs time to occur [24]. In this adiabatic condition, the strain rate in extension is 20 s$^{-1}$, where the kinetic requirement of SIC cannot be satisfied, leading to a partial crystallization and little influence on the mechanical behavior even in the satisfaction of low temperatures. As a result, the fast extension stress of NR exhibits almost temperature independence.

Due to the fast extension, the SIC mainly occurs in the keeping time of the stretched state. The SIC chain is longer than the amorphous chain [40]. The partially crystallized chain can relax the remaining chain and relax the stress (stress relaxation effect of SIC) [41,42]. The relaxed stress in the keeping time of stretched state, i.e. the difference of extension stress and retraction stress (indicated in Fig. 1 (a)), can thus be used to exhibit the SIC behavior. Specifically, its temperature dependence can exhibit the temperature dependence of SIC. In Fig. 2 (b), the relaxed stresses at different strains and different temperatures are compared. At almost every strain, the relaxed stress increases as temperature decreases. It indicates that the crystallinity increases as temperature decreases. Moreover, only at lower temperatures (0 °C, 10 °C), the relaxed stress becomes saturated from strain of 5. This may be due to the earlier onset strain of SIC as temperature decreases [43].

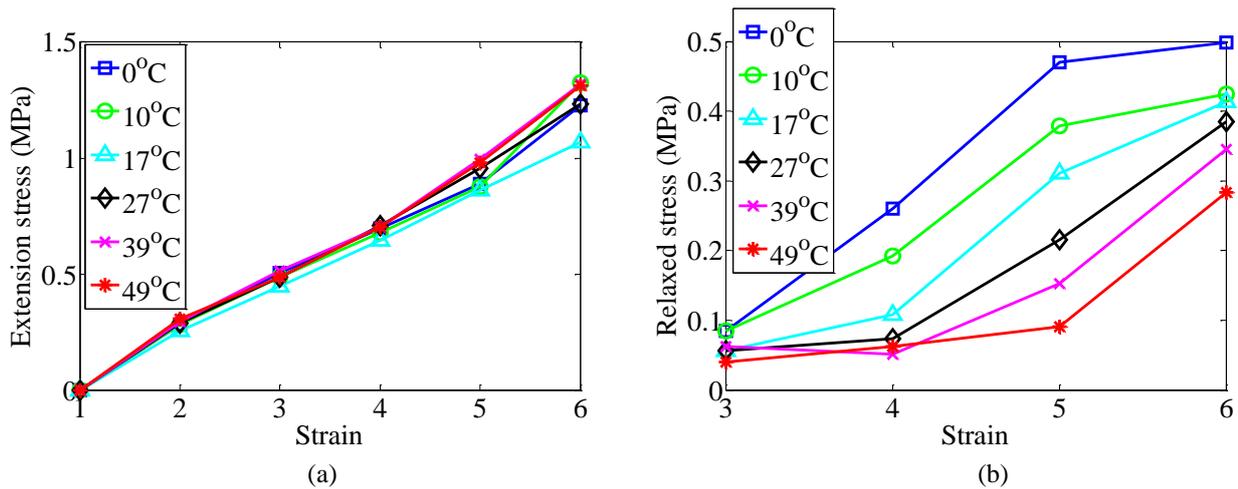

Fig. 2 (a) Extension stress and (b) relaxed stress at different temperatures in adiabatic condition.

Accompanying the stress, the adiabatic temperature changes (Δ$T$) are compared at different temperatures (Fig. 3). The upturn and hysteresis (not shown) of Δ$T$ prove that the Δ$T$ comes mainly from SIC phase transition [20,21,24,44]. In Fig. 3, from around strain of 3, Δ$T$ shows large temperature dependence. The onset strain of SIC is widely believed to be at around strain of 3 [22,38,45,46]. Thus, the SIC should be responsible for the temperature dependence of Δ$T$. The highest Δ$T$ is 12 K at strain of 6 and occurs at 10 °C. From 10 °C to 39 °C, the Δ$T$ decreases from 12 K to 5.8 K at strain of 6 (inset of Fig. 3), which is the same trend as the relaxed stress (Fig. 2 (b)). This is attributed to the decrease of crystallinity from 10 °C to 40 °C [36,37]. From 39 °C to 49 °C, the Δ$T$ shows a small variation with temperature. This is also consistent with the small crystallinity variation when temperature is higher than 40 °C [36].

It's surprising to see the sharp decrease of Δ$T$ from 10 °C to 0 °C. It is opposite to the variation trend of relaxed stress. When the temperature is close to 0 °C, the temperature-induced crystallization (TIC) in NR becomes prominent [47,48]. TIC is a slow process for un-stretched NR sample (in the timescale of hour) [47]. When the NR is stretched, the aligned chains may shorten the time needed for TIC and make it occur in the short deformation time (~30 s). The TIC occurred in the deformation process may be irreversible at around 0 °C. This can be shown by the yielding point of crystallinity at 8.5 °C (Fig. 4 (b)) [37]. Thus, the relaxed stress may come from both contribution of SIC and TIC, whereas the Δ$T$ only comes from the reversible SIC. Moreover, the TIC time may be still longer than adiabatic limit. Thus, it may not contribute to the measured Δ$T$ due to the heat loss with outer medium.

Whatever the responsible mechanism for temperature dependence of eC effect in NR (TIC and SIC), this measurement proves that the temperature span for eC effect of NR can be at least from 0 °C to 49 °C and the highest eC capacity occurs at around 10 °C. It proves the potential of NR for near room temperature cooling application.

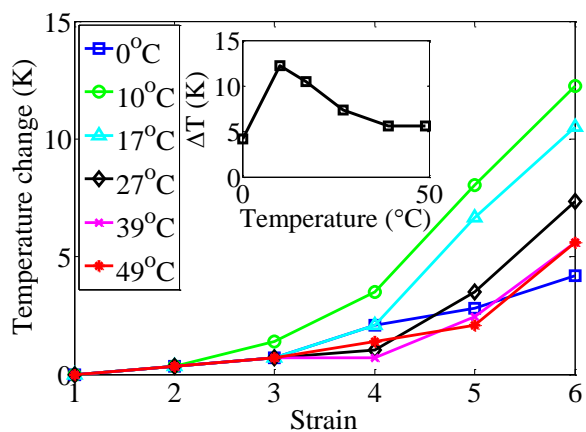

Fig. 3 Temperature dependence of adiabatic temperature changes (Δ*T*) at different strains. The inset shows this behavior at strain of 6.

In isothermal condition, the stress-strain curves at different temperatures are shown in Fig. 4 (a). The stress upturn is observed clearly at low temperatures (0 °C and 10 °C). The stress plateau in retraction can be observed at all tested temperatures and it is more obvious at low temperatures. Both the stress upturn and plateau are attributed to SIC [23,38,49]. Flory proposed two stress effects of SIC: one is the stress relaxation effect (discussed in Fig. 2 (b)), and the other is the stress hardening effect. The stress relaxation effect of SIC leads to the stress plateau in retraction. For stress hardening effect, SIC is a giant network point [50,51], which can increase the network chain density [23], leading to the stress upturn in extension [38].

Comparing the stress plateau regime of retraction with stress upturn regime of extension, the stress shows an opposite temperature dependence. In the stress upturn regime of extension, the higher crystallinity (Fig. 4 (b)) and larger stress (Fig. 4 (a)) at lower temperature are observed, which indicates the dominant stress hardening effect in extension. This large temperature dependence of extension stress is different from the fast extension stress in adiabatic condition (Fig. 2 (a)). In the slow isothermal extension, the kinetic requirement of SIC is satisfied. It can exhibit the temperature dependence and further contribute to the temperature dependence of extension stress. In retraction, a higher crystallinity (Fig. 4 (b)) and smaller stress (Fig. 4 (a)) are observed at lower temperature, which indicates the dominant stress relaxation effect of SIC. In conclusion, the opposite stress effects of SIC are believed to be responsible for the opposite temperature dependence of stress in stress upturn regime of extension and stress plateau regime of retraction.

The stress hysteresis is observed at all tested temperatures. The lower temperature induces higher crystallinity in both extension and retraction (Fig. 4 (b)). Coupled with the dominant stress hardening effect of SIC in extension and the dominant stress relaxation effect of SIC in retraction, it results in a larger hysteresis area at lower temperature (Fig. 4 (a)).

For shape memory alloys (SMAs), a different behavior is observed (Fig.4 (c)). In both extension and retraction, the stress plateau is observed. In both plateau regimes, the same temperature dependence of stress is observed, i.e. the higher temperature induces the larger stress. The stress hysteresis in SMAs is almost temperature independent. The different temperature dependent behaviors of NR and SMAs are attributed to the different types of phase transitions. For NR, it is the transition from amorphous state to crystalline phase, whereas for SMAs, it is the transition from one crystalline phase (austenite) to another crystalline phase (martensite).

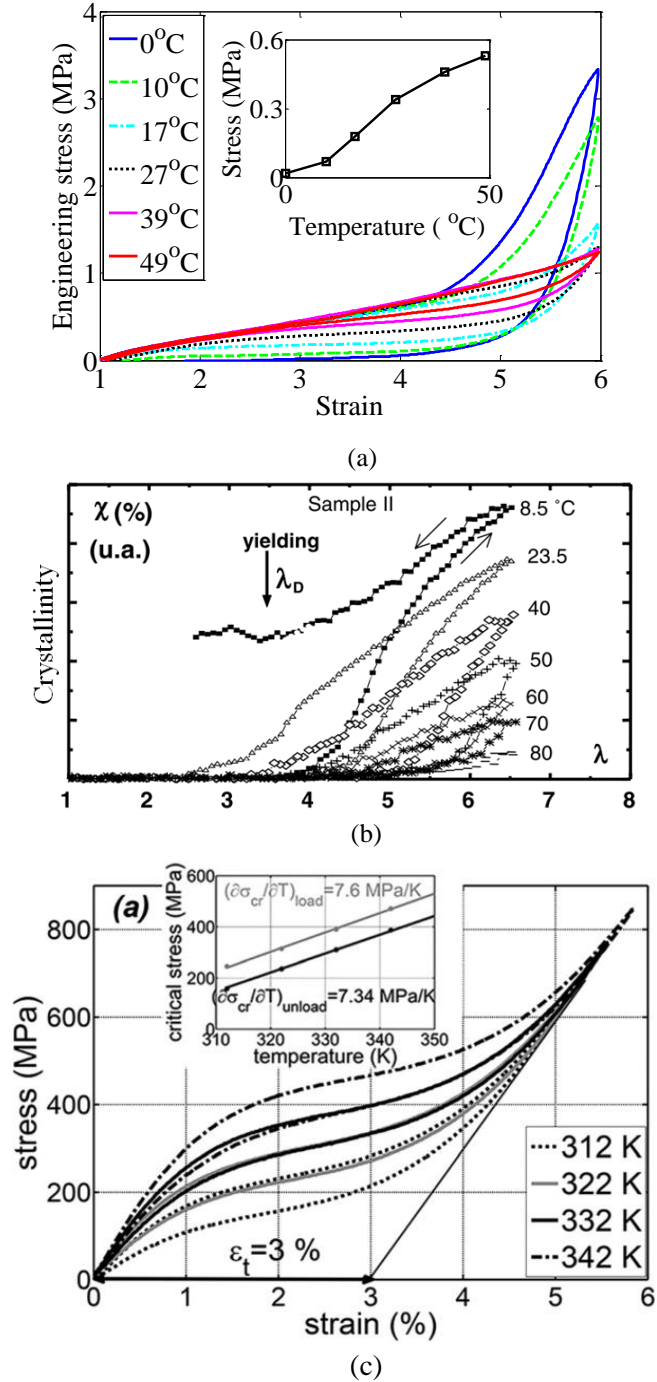

Fig. 4 (a) Temperature dependence of isothermal stress in NR. (b) Temperature dependence of crystallinity in NR [37]. *[Reprinted with permission from The European Physical Journal E Soft Matter, Chain orientation in natural rubber, Part I: The inverse yielding effect, 17, 2005, 247, P. A. Albouy, J. Marchal, and J. Rault, With permission of Springer]* (c) Temperature dependence of isothermal stress in SMA (Ni-Ti) [31]. *[Reprinted with permission from Journal of Applied Physics, 117, J. Tušek, K. Engelbrecht, L.P. Mikkelsen, and N. Pryds, Elastocaloric effect of Ni-Ti wire for application in a cooling device, 124901, Copyright [2015], AIP Publishing LLC]*

Most of the giant caloric materials undergo a first-order phase transition [52]. The direct calorimetric measurement for first-order phase transitions is using the differential scanning calorimetry (DSC). In addition, the isothermal entropy change $\Delta s$ can be obtained indirectly by using the temperature dependence of stress, as shown in the following formula:

$$\Delta s = \int_0^\varepsilon \left(\frac{\partial \sigma}{\partial T}\right)_\varepsilon d\varepsilon \tag{1}$$

The adiabatic temperature change $\Delta T$ can be deduced from the formula:

$$\Delta T = -\frac{T_0}{c} \cdot \Delta s \tag{2}$$

where $\sigma$ is the engineering stress, $\varepsilon$ is the engineering strain, $(\partial \sigma / \partial T)_\varepsilon$ is the Clausius-Clapeyron factor [53], $c = 1.8 \times 10^6 \, J \cdot K^{-1} \cdot m^{-3}$ is an estimation of specific volumetric heat of NR [54], and assumed to be independent of the temperature and strain.

In DSC measurement, the specimen is normally placed into a standard DSC alumina crucible [55]. In this set-up, the magnetic field or electric field is directly applied to the sample. Purpose-built DSC under different external fields have been developed to study the MC [56–58] and EC effects [59,60]. For measuring the eC effect of SMAs, the uniaxial stress is difficult to be applied. Fortunately, both the temperature-induced phase transition at zero stress and stress-induced phase transition are the same austenite–martensite transition. Thus, the direct measurement for the eC effect of SMAs by using DSC at zero stress can be used and shows a similar $\Delta s$ to the indirect measurement [12]. However, for eC effect of NR, the temperature-induced (at zero stress) and strain-induced phase transitions are different (TIC and SIC, respectively) for NR [29,61,62]. Thus, the direct measurement by using DSC at zero stress cannot be used for eC effect of NR. For measuring SIC, the needed stress is difficult to be applied in this set-up. Thus, the indirect measurement by using the Clausius-Clapeyron factor for the eC effect of NR becomes important. Moreover, it can provide a fast and easy measurement technique.

In case of SMAs, this indirect characterization leads to a reliable estimation of the entropy change [12,53]. As discussed before, a positive $(\partial \sigma / \partial T)_\varepsilon$ is obtained in both the stress plateaus of extension and retraction (Fig. 4 (c)). For NR, it is observed that $(\partial \sigma / \partial T)_\varepsilon > 0$ only in the stress plateau of retraction. The inset of Fig. 4 (a) shows the stress as a function of temperature in the stress plateau of retraction. As shown by the slope, the $(\partial \sigma / \partial T)_\varepsilon$ firstly increases and then decreases as temperature increases. This is different from the constant $(\partial \sigma / \partial T)_\varepsilon$ in Ni-Ti (inset of Fig. 4 (c)). For NR, the average value of $(\partial \sigma / \partial T)_\varepsilon$ is ~0.01 MPa/K from 0 °C to 49 °C, which is ~1/700 of that of Ni-Ti. By using Eq. (1), the $\Delta s$ of NR and Ni-Ti are ~0.04 MJ/ K.m$^3$ (44 J/K.kg) and ~0.22 MJ/K.m$^3$ (34 J/K.kg), respectively. Furthermore, the efficiencies of Ni-Ti and NR are calculated without considering the whole thermodynamic cycle [63]. In this case, only the extension work is considered as input mechanical work without energy recovery; and only the produced heat without heat recovery is considered as output heat. The results are shown in Table 1. The two eC materials show a similar mass energy density and efficiency. Compared with SMAs, the advantage of NR is its much lower stress. Considering the large strain and low volume energy density of NR, it is potential for a large-scale cooling application.

Table 1 Comparison of eC performance of Ni-Ti and NR

|  | **Ni-Ti [31]** | **NR** |
| --- | --- | --- |
| **Stress ($\sigma$)** | 800 MPa | 1.5 MPa (at 17 °C) |
| **Strain ($\varepsilon$)** | 0.06 | 6 |
| **Input mechanical work ($W = 1/2\sigma.\varepsilon$)** | 24 J/cm$^3$ | 4.5 J/cm$^3$ |
| **Isothermal entropy change ($\Delta s$)** | 0.22 J/K.cm$^3$; 34 J/K.kg | 0.04 J/K.cm$^3$; 44 J/K.kg |
| **Output heat ($Q = T. \Delta s$)** | 66 J/cm$^3$; 10 kJ/kg | 12 J/cm$^3$; 13 kJ/kg |
| **Efficiency ($\eta = Q/W$)** | 2.6 | 2.7 |

Furthermore, by using Eq. (2), the Δ$T$ is deduced and compared with the direct measurement at strain of 6 (Fig. 5). As temperature increases, the deduced Δ$T$ firstly increases and then decreases, which can agree reasonably with the direct measurement.

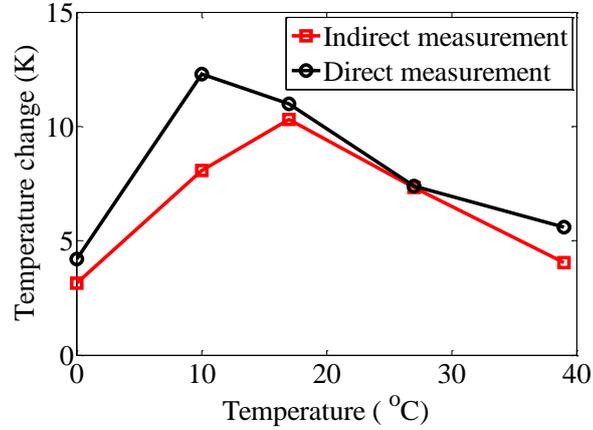

Fig. 5 Comparison of the direct measurement and indirect measurement at strain of 6.

In conclusion, the temperature dependence of elasocaloric (eC) effect of natural rubber (NR) is studied. Stress and temperature change are characterized in adiabatic and isothermal conditions. In adiabatic condition, the fast extension stress showed negligible temperature dependence from 0 $^o$C to 49 $^o$C. However, the relaxed stress, i.e. the difference between extension stress and retraction stress, decreases as the temperature increases. Relaxed stress is mainly due to strain-induced crystallization (SIC) and temperature-induced crystallization (TIC). The decrease of relaxed stress is due to the decrease of crystallinity as temperature increases. For adiabatic temperature change (Δ$T$), the maximum value is 12 K at strain of 6 and occurs at 10 $^o$C. From 10 $^o$C to 49 $^o$C, the Δ$T$ decreases from 12 K to 5.8 K. This is the same variation trend as the relaxed stress. It is attributed to the decrease of reversible SIC. From 10 $^o$C to 0 $^o$C, there is a sharp decrease of Δ$T$ from 12 K to 4 K. This is opposite to the variation trend of relaxed stress. It may be due to the slow and irreversible temperature-induced crystallization (TIC) when close to 0 $^o$C, which can contribute to the relaxed stress but cannot contribute to Δ$T$. Besides the intrinsic mechanism shown by relaxed stress and Δ$T$, this result proves that NR material has a large temperature span (at least from 0 $^o$C to 49 $^o$C) and it is potential for near room temperature cooling application.

In isothermal condition, both stresses in extension and retraction show large temperature dependence, which is different from the stress in adiabatic condition. For slow isothermal condition, the kinetic requirement of SIC is satisfied and its temperature dependent behavior can be revealed. Then, it can contribute to temperature dependence of stress. The retraction stress is close to the equilibrium state and gets a stress plateau regime, where Clausius-Clapeyron factor $(\partial \sigma / \partial T)_\varepsilon$ is positive. The deduced Δ$T$ from $(\partial \sigma / \partial T)_\varepsilon > 0$ can almost agree with the directly measured Δ$T$. These behaviors are totally different from those of shape memory alloys (SMAs). The eC performance of NR is compared with SMAs and its potential for large-scale application is shown.

The temperature dependence of SIC and TIC help understanding the temperature dependence of eC effect of NR. Accordingly, as NR is proved to be potential for a near room temperature and large-scale application, it will open the new research direction on SIC and TIC. For example, the caloric material needs a large Δ$T$. Higher crystallinity will contribute to larger Δ$T$. However, the maximum crystallinity of NR can only be up to ~20% [64]. Thus, there is much room to improve the crystallinity and thus the eC effect of NR.

**Acknowledgements**

The authors would like to thank China Scholarship Council (CSC).